# Collective excitations in twisted bilayer graphene close to the magic angle

Niels C.H. Hesp[1†], Iacopo Torre[1†], Daniel Rodan-Legrain[2†], Pietro Novelli[3,4†], Yuan Cao[2], Stephen Carr[5], Shiang Fang[5], Petr Stepanov[1], David Barcons-Ruiz[1], Hanan Herzig-Sheinfux[1], Kenji Watanabe[6], Takashi Taniguchi[6], Dmitri K. Efetov[1], Efthimios Kaxiras[5,7], Pablo Jarillo-Herrero[2*], Marco Polini[4*], Frank H.L. Koppens[1,8*]

[1]ICFO-Institut de Ciencies Fotoniques, The Barcelona Institute of Science and Technology, 08860 Castelldefels (Barcelona), Spain.

[2]Department of Physics, Massachusetts Institute of Technology, Cambridge, Massachusetts 02139, USA.

[3]NEST, Scuola Normale Superiore, I-56126 Pisa, Italy.

[4]Istituto Italiano di Tecnologia, Graphene Labs, Via Morego 30, I-16163 Genova, Italy.

[5]Department of Physics, Harvard University, Cambridge, Massachusetts 02138, USA.

[6]National Institute for Materials Science, Namiki 1-1, Tsukuba, Ibaraki 305-0044, Japan.

[7]John A. Paulson School of Engineering and Applied Sciences, Harvard University, Cambridge, Massachusetts 02138, USA.

[8]ICREA-Institució Catalana de Recerca i Estudis Avançats, 08010 Barcelona, Spain.

[†]Equally contributing authors

*To whom correspondence should be addressed: pjarillo@mit.edu, marco.polini@iit.it, frank.koppens@icfo.eu

**The electronic properties of twisted bilayer graphene (TBG) can be dramatically different from those of a single graphene layer, in particular when the two layers are rotated relative to each other by a small angle. TBG has recently attracted a great deal of interest, sparked by the discovery of correlated insulating and superconducting states, for twist angle $\theta$ close to a so-called "magic angle" $\approx \mathbf{1.1°}$. In this work, we unveil, via near-field optical microscopy, a collective plasmon mode in charge-neutral TBG near the magic angle, which is dramatically different from the ordinary single-layer graphene intraband plasmon. In selected regions of our samples, we find a gapped collective mode with linear dispersion, akin to the bulk magnetoplasmons of a two-dimensional (2D) electron gas. We interpret these as interband plasmons and associate those with the optical transitions between quasi-localized states originating from the moiré superlattice. Surprisingly, we find a higher plasmon group velocity than expected, which implies an enhanced strength of the corresponding optical transition. This points to a weaker interlayer coupling in the AA regions. These intriguing optical properties offer new insights, complementary to other techniques, on the carrier dynamics in this novel quantum electron system.**

When two layers of graphene are superimposed with a small twist angle $\theta$, they form a triangular moiré lattice with a lattice constant $d$ that is related to $\theta$ by $d = d_0/[2\sin(\theta/2)]$, $d_0 \approx 0.246$ nm being the lattice constant of single-layer graphene[1–4]. A top view of TBG reveals regions where the two sheets are locally in the AA-stacking configuration surrounded by regions where the stacking configuration is of the more energetically favoured AB- or BA-type (Bernal stacking)[5].

Electrons can tunnel from one layer to the other with an amplitude that depends on the *local* alignment between the two layers[1,3]. The interlayer tunnelling amplitude is therefore spatially modulated with the periodicity of the moiré lattice. Effectively, this produces a background scalar potential and non-Abelian gauge field, acting on the graphene Dirac fermions[6]. These two potentials (with an amplitude on the order of $100$ meV) localize electronic states close to the charge neutrality point (CNP) in the regions where the alignment between the two layers is AA–like[7–10]. In a band structure picture, these two contributions yield a pair of nearly-flat bands close to CNP at the magic angle, which, due to their high density of states, are held responsible for the observed correlated phenomena[9–18] [19]. Switching off the scalar potential enhances the flatness of the bands making them perfectly flat throughout the superlattice Brillouin zone (BZ) at the magic angle[6,20].

Several experimental probes have been used to explore the physics of TBG, including electronic transport[9–14,21], quantum capacitance[15], scanning tunneling microscopy[16–19] and scanning magnetometry[22], also unveiling similar phenomena in twisted double bilayer graphene[23–25] and trilayer graphene on hexagonal boron nitride (hBN)[26]. However, all these techniques are sensitive only to the static (very low frequency) response of the system. In systems where electron-electron (e-e) interactions play a dominant role, experimental techniques that probe the response to perturbations carrying a finite in-plane wavevector $q$ and angular frequency $\omega$ are expected to be rich sources of information. One of these techniques is scattering-type scanning near-field optical microscopy (s-SNOM)[27–31], which enables the measurement of the dispersion relation of collective electronic excitations, such as Dirac plasmons in doped graphene[27,28,32].

**Optical properties of TBG**

The order of magnitude of the energy separation between the nearly-flat bands and the nearest conduction and valence bands is $\approx 100$ meV for $\theta$ close to the magic angle. This justifies our interest in the optical properties and collective excitations[33–35], as probed by s-SNOM (Fig. 1a-b), in the mid-infrared (MIR) region of the electromagnetic spectrum where photons have energies $\hbar\omega$ in the range $80 - 200$ meV, i.e. comparable to the above-mentioned energy scale. These energies are on the other hand much larger than the energy separation between the pair of nearly-flat bands, considered in Ref. 36. The square modulus of the wave functions of one of the nearly-flat bands and of the first conduction band, evaluated at the $K$ point of the Brillouin zone (BZ) (close to which most of the relevant transition occur), are visualized in Fig. 1c. The optical transitions relevant for this work are shown in Fig. 1d.

We can qualitatively understand the optical properties of this system in the following way. When light impinges on TBG, its time-periodic electric field shakes electrons around their equilibrium positions (the AA sites, forming a triangular lattice) or—in the more rigorous language of band theory—it induces an interband transition (Fig. 1d). If the field carries a finite in-plane wavevector $q$, the shaking electrons will build up an oscillating charge density with the same wavevector (plus harmonics due to exchange of reciprocal lattice vectors). This oscillating charge density, in turn, creates an oscillating electric field that adds to the external field. If $q$ and $\omega$ are correctly matched, this induced field can be strong enough to sustain the oscillation even after the external field has been turned off. This resonant behaviour gives rise to collective modes that are called interband plasmons[37] (schematically depicted in Fig. 1e). At $q = 0$, this collective excitation has the same frequency as the bare interband transition. Upon increasing $q$, as we show in this work, it acquires a finite dispersion. Thus, these excitations do propagate with a finite group velocity, akin to graphene Dirac plasmons[27,28,38]. The dispersion depends on the degree of band nesting, which is the phenomenon of two bands being parallel in energy-wavevector space, and details of the e-e interaction potential, which is heavily influenced by screening from nearby dielectrics. We note that a similar collective mode occurs between the Landau-level flatbands of a two-dimensional (2D) parabolic-band electron gas in a perpendicular magnetic field. In this case, while single electrons oscillate at the cyclotron frequency $\Omega_c$, e-e interactions induce a collective mode, known as a bulk magnetoplasmon, which, at long wavelength, has a linear dispersion[39] $\omega(q) = \Omega_c + sq$ with group velocity $s > 0$ reflecting its propagating character.

**Near-field experiments on TBG**

We now turn to a description of our experimental findings. We fabricate TBG samples using the tear-and-stack method[40,41]. These are encapsulated in hBN, and placed on a metal gate (see Sect. 1 of S.I.). The twist angle $\theta$ is determined from cryogenic transport measurements[40] (see Sect. 2 of S.I.). We then perform s-SNOM measurements with MIR light (free-space wavelength $\lambda_0$ in the range $5 - 11$ μm) in ambient conditions ($T = 300$ K). We generate a nanoscale light hotspot by focussing a laser beam on the apex of a sharp (apex radius $\approx 25$ nm) metallic atomic force microscope (AFM) tip (Fig. 1a). This hotspot interacts with the charge carriers and produces collective excitations that are reflected by interfaces, return to the tip, and are finally converted into a scattered field, which is measured by a photodetector. By scanning the tip position, we acquire, simultaneously, a spatial map of the backscattered light intensity $S_{\text{opt}}$ and AFM topography. Noise and far-field contributions to the optical signal are strongly reduced by locking to the third harmonic of the tapping frequency of the tip. The spatial resolution of the obtained images is limited only by the tip radius[42], see also Sect. 3 of S.I..

Figure 1b shows a typical near-field image of TBG with no gate voltage applied (at zero applied voltage the TBG is close to charge neutrality, see Sect. 1-2 of S.I.) and a twist angle $\theta = 1.35°$. The most evident feature is the presence of well-defined optically active areas where $S_{\text{opt}}$ displays an oscillatory spatial behaviour. The latter has a characteristic period $\approx 80$ nm, about one order of magnitude larger than $d$. We attribute this oscillatory behaviour to the excitation of a propagating collective electronic mode, as schematically illustrated in Fig. 1e. The fact that we observe these interference patterns in *ungated* TBG is in stark contrast with the *intraband* collective electronic excitations (Dirac plasmons) of single-layer and bilayer graphene, where high doping levels (above $10^{13}$ cm$^{-2}$) are required to propagate at the frequencies we focus on ($\hbar\omega \sim 200$ meV)[29,32,38].

The areas where the collective excitation is visible partially correlate with tiny changes in the sample topography, as discussed in Sect. 4 of S.I. On few selected samples we applied the so-called AFM-brooming technique[43,44], in order to remove residues from the surface. This consists in sweeping repeatedly the sample with an AFM tip in contact mode, while applying a quite large force $\approx 30$ nN, corresponding to a pressure of $\approx 30$ MPa. We see that AFM-brooming is able to change the position and size of the optically active areas (see Sect. 4 of S.I.). This suggests that the local strain distribution plays an important role in the existence of optically active areas. As can be seen from Fig. 1b, the boundaries of the optically active areas are typically formed by sequences of arcs with radii ≈ 120 − 200 nm. This shape may stem from a boundary between two structural phases. Similar features are present in 5 out of 13 analysed devices, including a twisted double bilayer graphene device. A complete list of the analysed devices is reported in Sect. 1 of S.I.. In the following we focus on the $\theta = 1.35°$ device depicted in Fig. 1b, for which the features are most pronounced.

To get more insight into the nature of the collective excitations we probe their frequency dependence by repeating the near-field measurements at different excitation energies, changing $\lambda_0 = 2\pi c/\omega$. Figure 2a-d shows a dramatic change in the interference pattern for small variations in $\lambda_0$, while the boundaries of the areas where the sample is optically active remain at a fixed position. These data show the dispersive character of the propagating collective excitations that move in Fabry-Pérot-like cavities, due to reflecting interfaces[45,46].

To be more quantitative, we extract one-dimensional (1D) cuts of the measured $S_{\text{opt}}$ along two specific lines (see arrows in Fig. 2d). The resulting 1D profiles are shown in Fig. 2e as lines for a few representative photon energies, while a colormap as a function of tip position and frequency is reported in Fig. 3a. The oscillating signal is well fitted by the following expression, representing a tip-launched, tip-detected wave reflected at an interface: $S_{\text{opt}}(x) = \text{Re}[A\, x^{-1/2}\, e^{2iqx}] + Bx$. Here, $x$ is the tip position along the line cut, as measured from the interface, $A \equiv A_1 + iA_2$ and $q \equiv q_1 + iq_2$ are complex fit parameters, and $B$ represents a linear background[32]. Note the factor of two in the exponential function that appears because the collective excitation makes a full round trip between the tip and the reflecting interface. Our fitting procedure yields quantitative results for the real part $q_1$ of the wavevector $q$ while the imaginary part $q_2$ has a larger error (see Sect. 5 of S.I. for further details).

From the extracted values of $q_1(\omega)$ we can construct a dispersion curve for the collective excitation as shown in Fig. 3b. For energies above 200 meV, the dispersion is approximatively linear with a group velocity $s \approx 1.3 \cdot 10^6$ m/s, and crosses the $q_1 = 0$ point for $\hbar\Omega_{\text{exp}} \approx 190$ meV. For lower energies, the typical discretization pattern of a finite size cavity appears (i.e. where the distance between the reflecting interfaces is comparable to the plasmon wavelength $2\pi/q_1$). Remarkably, the group velocity is larger than theoretically anticipated. See e.g. Ref. 37, where flat plasmonic bands were predicted, and the discussion below. As we will see, this points to a larger spectral weight in the optical transitions. Clearly, the observed nearly-linear dispersion, initiating from a finite energy $\hbar\Omega_{\text{exp}}$ for $q_1 = 0$, is very different from the typical Dirac plasmon dispersion of doped graphene (see Sect. 6 of S.I. for more detailed comparisons). Instead, the observed linear dispersion resembles more the one of a bulk magnetoplasmon[39]: $\omega(q) = \Omega_{\text{exp}} + sq_1$.

**Extraction of the optical conductivity**

To relate our observations to the electronic bands in the moiré superlattice, we extract the value of the optical conductivity $\sigma(\omega)$ for the optically active regions. In the local approximation[47] (i.e. where the optical conductivity is taken to be independent of $q$ and contributions from reciprocal lattice vectors $\boldsymbol{G} \neq \boldsymbol{0}$ are neglected), the longitudinal dielectric function[48] is given by $\epsilon(q,\omega) = 1 + iq^2 V_{q,\omega}\, \sigma(\omega)/\omega$, where $V_{q,\omega} = 2\pi F(q,\omega)/[\tilde{\epsilon}(\omega)q]$ is the 2D Fourier transform of the Coulomb potential[49], the permittivity $\tilde{\epsilon}(\omega) = \sqrt{\epsilon_\parallel(\omega)\epsilon_\perp(\omega)}$ takes care of the optical response at frequency $\omega$ of the hBN crystal slabs[50] surrounding the TBG sample, and $F(q,\omega)$ is a form factor that takes into account the finite thickness of the hBN slabs (see Sect. 7 of S.I.). Finite thickness effects are important close to the upper edge of the hBN reststrahlen band where the in-plane permittivity $\epsilon_\parallel(\omega)$ vanishes and the out-of-plane decay length of the mode diverges. Neglecting the finite thickness of hBN would lead to a wrong dispersion relation, yielding a collective mode that does not enter the upper reststrahlen band (see Fig. 3b). Collective modes can be found by solving[48] $\epsilon(q,\omega) = 0$, or by looking at the peaks of the loss function $L(q,\omega) = -\text{Im}[\epsilon(q,\omega)^{-1}]$. From the measured collective excitation dispersion, we can find the imaginary part $\sigma_2(\omega)$ of the local conductivity, using the expression $\sigma(\omega) = i\omega/(q^2 V_{q,\omega})$ and neglecting the imaginary part of $q$. The results are shown in Fig. 3c.

The simplest possible fitting function, of the Drude form $\sigma_2(\omega) = G_0 W_0/(\hbar\omega)$—where $G_0 = 2e^2/h$ is the conductance quantum and $W_0$ is a fitting parameter with dimensions of energy—yields $W_0 \approx 1100$ meV (that would correspond, for two uncoupled single-layer graphene sheets, to a Fermi energy $\epsilon_F \approx 550$ meV in each layer) and a very poor fit. This confirms that our data are not consistent with a regular intraband graphene Dirac plasmon. A much better fit is obtained by using the following resonant form $\sigma_2(\omega) = G_0 W_\text{exp} \hbar\omega/(\hbar^2\omega^2 - \hbar^2\Omega_\text{exp}^2)$, with $W_\text{exp}$ and $\Omega_\text{exp}$ fitting parameters. We find $\hbar\Omega_\text{exp} \approx 180$ meV and a spectral weight $W_\text{exp} \approx 300$ meV for both presented datasets.

**Theory of interband transitions in TBG**

We now seek a theoretical justification for the resonant lineshape extracted from the experimental data and for the values we have found for $\Omega_\text{exp}$ and $W_\text{exp}$. At the level of the random phase approximation (RPA) for the dynamical dielectric function[48] $\epsilon(q,\omega)$, the unknown quantity $\sigma(\omega)$ is approximated by using its value for the non-interacting 2D electron system in TBG. The latter can be calculated exactly by employing the Kubo formula[48], once the eigenstates $|\mathbf{k},\nu\rangle$ and bands $\epsilon_{\mathbf{k},\nu}$ of the single-particle problem are given. The quantities $|\mathbf{k},\nu\rangle$ and $\epsilon_{\mathbf{k},\nu}$ can be found from a band structure calculation for TBG at a given $\theta$. Here, we have used results obtained from *ab initio* $k \cdot p$ perturbation theory[51], which accurately accounts for the effects of intrinsic atomic relaxation in *pristine* samples. The resulting bands $\epsilon_{\mathbf{k},\nu}$ are shown in Fig. 4a. We clearly see that band nesting occurs near the $K$ point of the superlattice BZ, where two (relatively flat) bands—say $\nu$ and $\nu'$, connected by vertical lines with arrows in Fig. 4a—are such that $\nabla_\mathbf{k}\epsilon_{\mathbf{k},\nu} \simeq \nabla_\mathbf{k}\epsilon_{\mathbf{k},\nu'}$, in a range of values of $\mathbf{k}$. In other words, the bands are parallel to each other for a wide range of $\mathbf{k}$. The joint density of states for these pairs of bands is large at the transition frequency and the resultant optical absorption spectrum $\sigma_1(\omega) \equiv \text{Re}[\sigma(\omega)]$ has a peak at a near frequency $\Omega_\text{th}$ as shown in Fig. 4b. At the CNP, $\theta = 1.35°$, and $T = 300$ K (used for all calculations in this work), we find $\hbar\Omega_\text{th} \approx 115$ meV and an associated spectral weight $W_\text{th} \equiv 2\hbar \int_\text{peak} d\omega\, \sigma_1(\omega)/(\pi G_0) \approx 64$ meV. Because of causality, $\sigma_1(\omega)$ and $\sigma_2(\omega)$ are related by a Kramers-Kronig transform[48]. The resonant lineshape introduced above for $\sigma_2(\omega)$ yields $\sigma_1(\omega) = \pi G_0 W_\text{exp}[\delta(\hbar\omega - \hbar\Omega_\text{exp}) + \delta(\hbar\omega + \hbar\Omega_\text{exp})]/2$. This implies that our simple resonant fitting formula for $\sigma_2(\omega)$ represents the peak seen in the microscopically calculated $\sigma_1(\omega)$ at $\Omega_\text{th}$—see Fig. 4b—with a delta peak at $\Omega_\text{exp}$ with spectral weight $W_\text{exp}$.

While $\Omega_\text{th}$ is in a reasonable agreement with $\Omega_\text{exp}$, there is significant disagreement with the spectral weight since $W_\text{th} \ll W_\text{exp}$. The sources of this spectral weight mismatch can be multiple. To gain understanding, we resort to a more flexible continuum band-structure model[52]. This contains two parameters, $u_0$ and $u_1$, denoting the inter-layer coupling in the AA regions and AB&BA regions, respectively. Results based on such continuum model with the choice[52] $u_0 = 79.7$ meV and $u_1 = 97.5$ meV present only minor quantitative differences with respect to those of *ab initio* $k \cdot p$ perturbation theory shown in Fig. 4b and are reported in Sect. 8 of S.I.. Also, calculations in Ref. 37 give qualitatively similar results. The same continuum model with $u_0 = 0$ and $u_1 \neq 0$, which is endowed with unitary particle-hole symmetry, has been introduced in Ref. 20 as an idealization of reality.

Intriguingly, we find that the conductivity calculated from this chirally-symmetric continuum model (CS-CM)[20]—for $u_0 = 0$ and $u_1 = 97.5$ meV—displays a much better agreement with our experimental data. The bands $\epsilon_{k,\nu}$ of the CS-CM are shown in Fig. 4c, while the optical absorption spectrum is reported in Fig. 4d. Also here, $\sigma_2$ displays a resonant profile, but the resonant energy is $\hbar\Omega_{\text{th}} \approx 199$ meV and the spectral weight is $W_{\text{th}} \approx 162$ meV, in much better agreement with our experimental results. The loss function, calculated from this CS-CM is shown in Fig. 3b and overlaps very well with the superimposed experimental data. We therefore find that the optical spectral weight is strongly enhanced in the optically active regions and this enhancement can be explained by an effective suppression of the AA interlayer coupling in the same regions.

Motivated by this finding, we perform a systematic scan of the AA tunnelling amplitude in a range going from $u_0 = 0$ meV —corresponding to the CS-CM—to $u_0 = 79.7$ meV, i.e. the value given in Ref. 52. For each value of $u_0$ we calculate the band structure, extract the optical conductivity, and fit it with a resonant profile to extract the parameters $\hbar\Omega_{\text{th}}$ and $W_{\text{th}}$. The results are shown in Fig. 5. The resonant frequency increases monotonically with decreasing $u_0$ and crosses the experimentally measured value around $u_0 \approx 40$ meV. The spectral weight has instead a non-monotonic behaviour but gets closest to the experimental data approximately in the range $20 \text{ meV} < u_0 < 40$ meV. We performed the same procedure on the *ab initio* $k \cdot p$ perturbation theory model by scaling the parameters corresponding to the AA tunnelling. The results are qualitatively similar, apart from a rigid shift, as shown in Fig. 5. The rigid shift is primarily caused by a small difference between the models' effective AB coupling $u_1$. This can also be viewed as a tunable parameter of the theory and controls the location of the magic-angle.

An extensive theoretical discussion of collective modes in TBG as a function of the band-structure parameters can be found in Ref. 53.

Our experiments therefore suggest that, in particular regions of the sample, the AA tunnelling amplitude is significantly reduced with respect to the AB tunnelling amplitude but still has a non-vanishing value. This finding is compatible with the results of Refs. 9,10,40 that reported on the gap size between the flat bands and the first excited band at the Γ point of the BZ. This quantity provides a direct measure of the difference $u_1 - u_0$ and was found to be in the range $30 - 60$ meV.

The apparent suppression of tunnelling in the AA regions compared to the one in the AB&BA regions (i.e. the fact that $u_0 < u_1$) can stem from e-e interactions or extrinsic effects. It is known[6,20] that $u_1$ is responsible for a (non-Abelian) gauge field acting on the electron system, while $u_0$ induces a scalar potential. Electron-electron interactions act between density fluctuations and therefore, because of the continuity equation, between longitudinal current fluctuations. Screening due to e-e interactions will therefore tend to suppress the longitudinal field due to $u_0$, while having a smaller impact on the transverse gauge field due to $u_1$.

Extrinsic factors can also alter, locally, the value of $u_0$. We suspect that these include the way samples are prepared, the aforementioned AFM-brooming procedure, and, possibly, the hBN encapsulation. It frequently happens that samples prepared in different laboratories display some macroscopic differences in their physical characteristics, such as twist angle[22], electrical transport[9–12,14,21] and spectroscopic features[16–19].

Before concluding, one may also hypothesize that our samples present a highly inhomogeneous strain distribution with patches where the associated pseudo-magnetic field $\boldsymbol{B}_S = B_S \hat{\boldsymbol{z}}$ is finite and nearly uniform[54] and regions where $\boldsymbol{B}_S = \boldsymbol{0}$. A resonant conductivity profile, as observed in experiments, would naturally arise in this case at the frequency of the pseudo-cyclotron resonance. We analyse this potential explanation of our observations in Sect. 9 of S.I. and conclude that it is unlikely since it would require an unreasonably large amount of strain to match the observed resonant frequency.

**Conclusions**

In summary, we have observed dispersive interband collective excitations in TBG close to the magic angle, with larger-than-expected group velocity and thus a larger spectral weight of the infrared optical transitions. The usefulness of models with reduced AA tunnel coupling[20,51,52] in interpreting our experimental data could point to the enhanced role of e-e interactions. A deep understanding of collective excitations in TBG may also shed light on superconducting states. Indeed, it has been recently suggested[55] that collective modes in TBG can mediate pairing with $T_c \sim 10$ K. These calculations were restricted to dynamical screening stemming from the nearly-flat bands close to the CNP, and included transitions to higher excited bands by virtue of an effective renormalized dielectric constant. It will therefore be very interesting to generalize the Migdal-Eliashberg theory of 'plasmonic' superconductivity[55] to include our interband plasmon. Future low-temperature studies can further elucidate the role of electronic correlations in the upper bands, while terahertz near-field imaging can offer a local probe of the electronic phase transitions.

**Acknowledgements**

We thank F. Vialla for providing the illustration in Fig. 1a. We acknowledge A.H. MacDonald, F. Guinea, F. Mauri, R.K. Kumar, and A. Tomadin for useful discussions.

The research leading to these results has received funding from the European Union's Horizon 2020 research and innovation program under grant agreement No. 785219 Graphene Flagship. This work was also partially supported by the ERC TOPONANOP under grant agreement No. 726001 and the MINECO Plan Nacional Grant 2D-NANOTOP under reference No. FIS2016-81044-P. N.C.H.H., I.T., P.S., D.B.R., H.H.S., D.K.E. and F.H.L.K. acknowledge financial support from the Spanish Ministry of Economy and Competitiveness through the 'Severo Ochoa' program for Centres of Excellence in R&D (SEV-2015-0522), from Fundació Privada Cellex and Fundació Privada Mir-Puig, and from Generalitat de Catalunya through the CERCA program. N.C.H.H. acknowledges funding from the European Union's Horizon 2020 research and innovation programme under the Marie Skłodowska-Curie grant agreement Ref. 665884. D.B.R. acknowledges funding from the "Secretaria d'Universitats i Recerca de la Generalitat de Catalunya i del Fons Social Europeu". H.H.S. acknowledges funding under the Marie Skłodowska-Curie grant agreement Ref. 843830. F.H.L.K. acknowledges the Mineco grants Ramón y Cajal (RYC-2012-12281, and the Agency for Management of University and Research Grants (AGAUR) 2017 SGR 1656. D.K.E. acknowledges support from the H2020 Programme under grant agreement No. 820378, Project: 2D·SIPC and the La Caixa Foundation. P.N. and M.P. have been supported by the European Union's Horizon 2020 research and innovation program under Grant Agreement No. 785219, GrapheneCore2. Work at MIT has been primarily supported by the National Science Foundation (award DMR-1809802), the Center for Integrated Quantum Materials under NSF grant DMR-1231319, and the Gordon and Betty Moore Foundation's EPiQS Initiative through Grant GBMF4541 to P.J.H. for device fabrication, transport measurements, and data analysis. This work was performed in part at the Harvard University Center for Nanoscale Systems (CNS), a member of the National Nanotechnology Coordinated Infrastructure Network (NNCI), which is supported by the National Science Foundation under NSF ECCS award No. 1541959. D.R.L. acknowledges partial support from Fundacio Bancaria 'la Caixa' (LCF/BQ/AN15/10380011). S.F. was supported by the STC Center for Integrated Quantum Materials, NSF Grant No. DMR1231319. S.C. and E.K. were supported by ARO MURI Award W911NF-14-0247. K.W. and T.T. acknowledge support from the Elemental Strategy Initiative conducted by the MEXT, Japan and the CREST (JPMJCR15F3), JST.

**Author contributions**

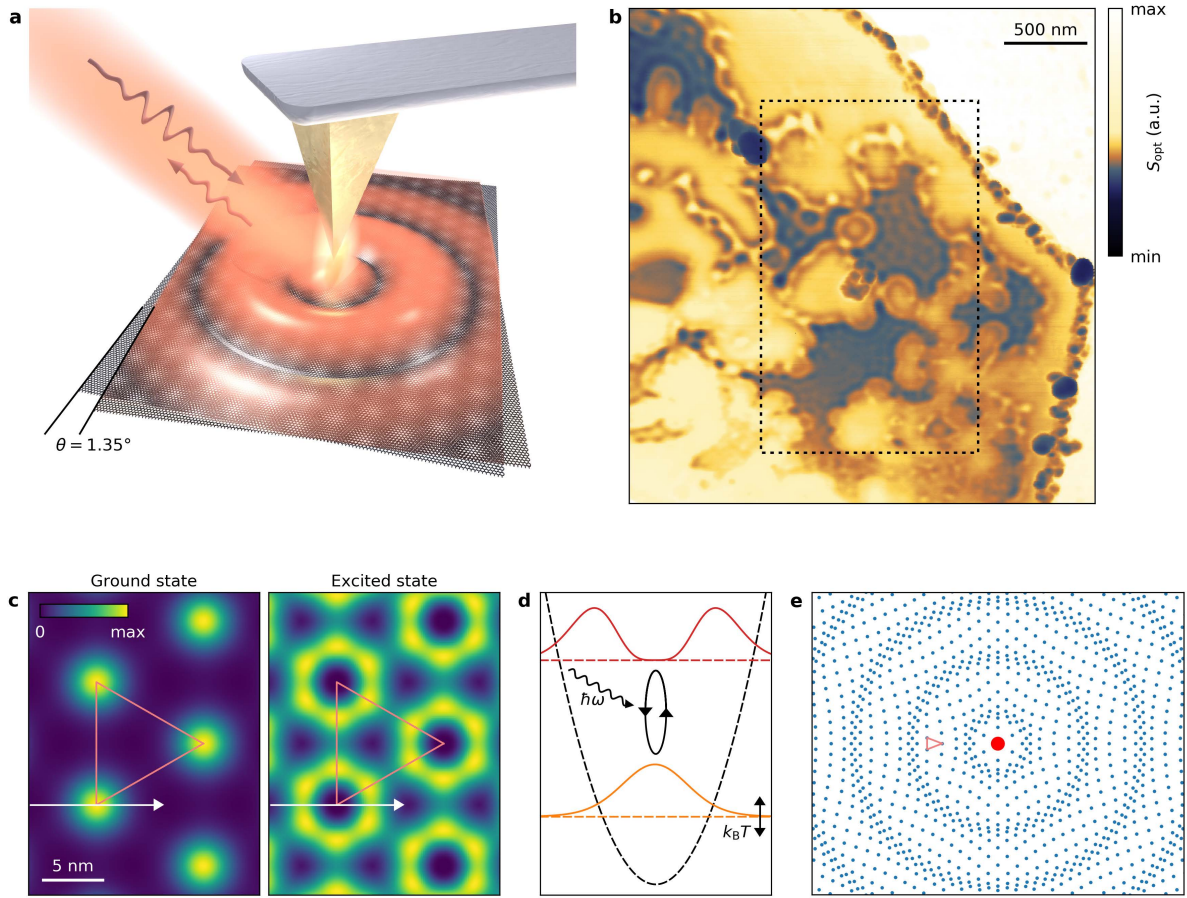

**Figure 1 | Collective excitations in twisted bilayer graphene. a**. Illustration of the scattering-type scanning near-field microscopy experiment: an AFM metallic tip is illuminated by infrared light, which provides sufficient momentum to launch a collective excitation in twisted bilayer graphene. The plasmon can, in turn (e.g. by reflection from an edge or interface), scatter into light at the tip. This scattered light is detected by a photodetector. The curved arrows indicate the light impinging on the tip (coming from the laser) as well as the scattered light (going to the detector). **b**. Image of the near-field amplitude obtained by scanning the AFM tip and recording the photodetector signal. Propagating collective excitations are visible in certain areas as periodic interference fringes. The illumination photon energy is $\hbar\omega = 219$ meV. **c**. Square modulus of the wave function $|\langle r|k=K,v\rangle|^2$, associated with one of the flat bands (left) and the first excited band (right), evaluated at the $K$ point of the moiré superlattice Brillouin zone. These states are mostly localized around the regions with local AA-stacking (which form a triangular lattice marked by the red triangle) and are involved in the relevant optical transitions. **d**. Line cuts of the wave functions along the white arrows in panel **c** represented in a harmonic confinement potential[6] (black dashed line), with the coloured horizontal dashed lines indicating the energies of the states. An interband transition occurs between the lower-energy state and the excited state. A similar transition is happening, for holes, between the corresponding pair of states, approximately related to the illustrated ones by electron-hole symmetry. **e**. Simplified and classical representation of the collective movement induced by e-e interactions of an ensemble of carriers in the moiré superlattice (red triangle corresponding to those in panel **c**), with the red dot marking the point of excitation. The magnitude of the movement is enlarged for clarity.

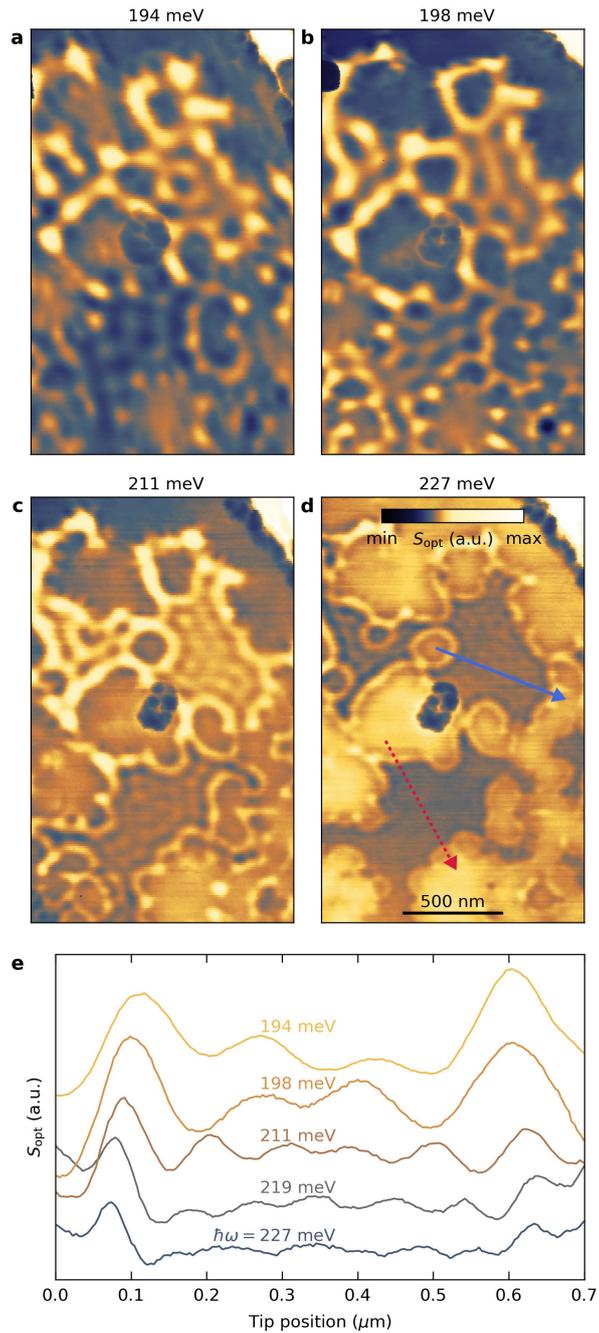

**Figure 2 | Controlling the wavelength of interband plasmons. a-d**. Near-field amplitude images at different excitation energies $\hbar\omega$ of the area marked in Fig. 1**b**. Solid and dashed arrows in panel **d** indicate linetraces associated to data in panel **e** and Fig. 3**a**. **e**. Linetraces along the solid blue arrow in panel **d**, visualizing the strong dependence of the plasmon wavelength on the excitation energy. Lines are vertically separated for clarity.

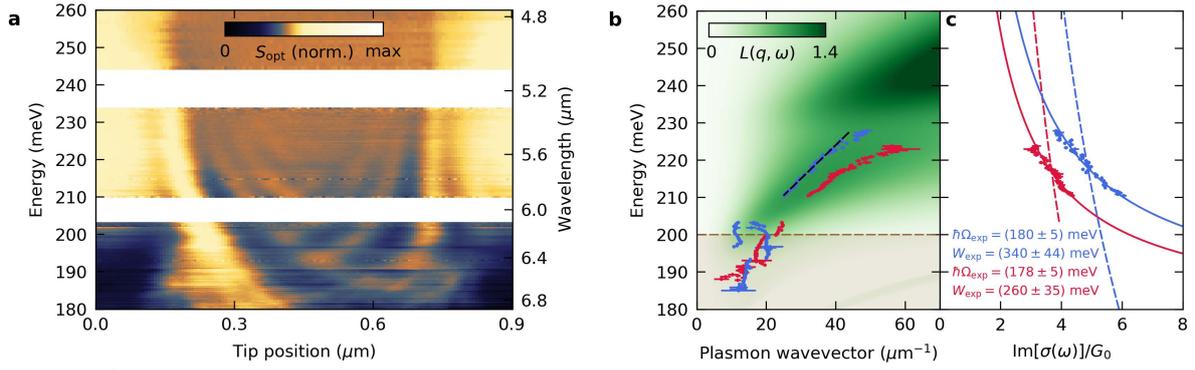

**Figure 3 | Extracting the optical conductivity from the plasmon dispersion. a**. Near-field amplitude along the dashed red arrow in Fig. 2**d**, for a range of excitation energies. To highlight the plasmonic modes, we normalize each line to the average near-field amplitude within the reflecting interfaces. The white gaps are gaps in the spectrum of the excitation laser. **b**. Dispersion relation $q_1(\omega)$ determined from fitting individual linetraces in panel **a** to a sinusoidal function (red points). The blue points are obtained in a similar way but from a slightly different location (solid blue arrow in Fig. 2**d**). We extract the plasmon group velocity (black dashed line) from a linear fit on the blue data points. The horizontal dashed line marks the threshold of the hBN reststrahlen band. The colorplot represents the loss function, calculated from the chirally-symmetric continuum model[20]. **c**. Extracted values of the optical conductivity with the same colour coding as in panel **b**. Dots represent experimental data, dashed lines are Drude fits, while solid lines are fits with resonant profiles.

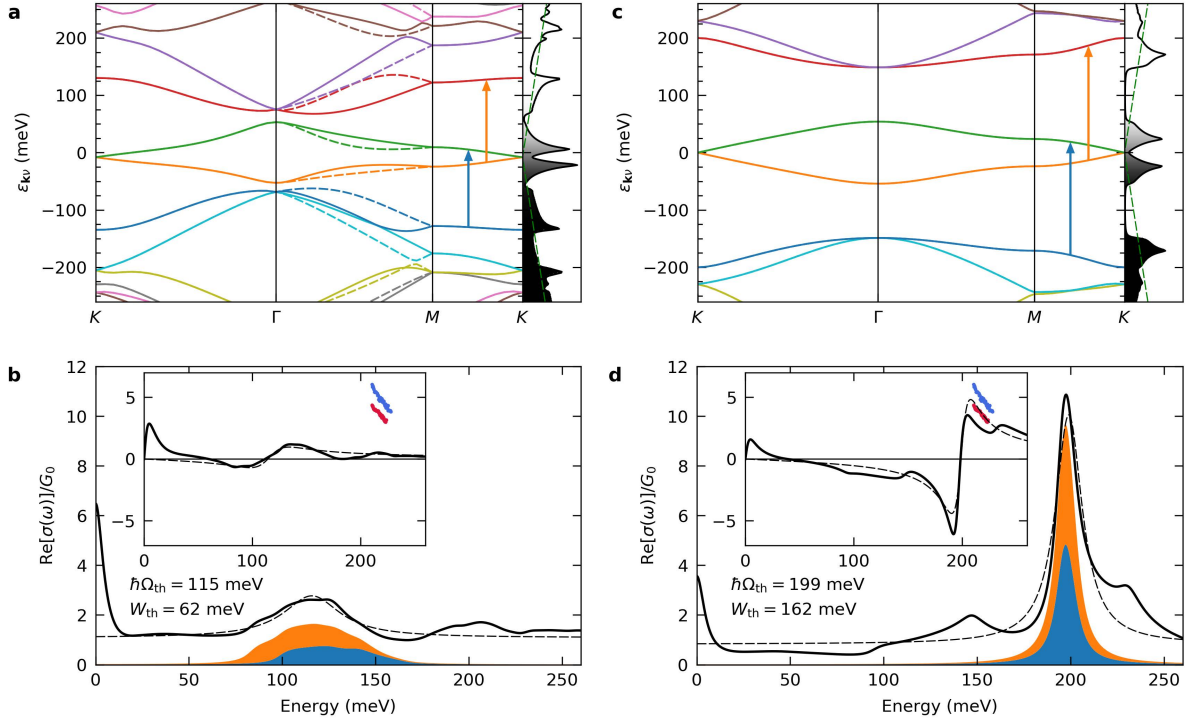

**Figure 4 | Electronic band structure and optical conductivity of twisted bilayer graphene with $\theta = 1.35°$. a.** Electronic band structure $\epsilon_{k\nu}$ of TBG with $\theta = 1.35°$ along the $K\Gamma MK$ contour of the superlattice Brilloun zone from *ab initio $k \cdot p$* perturbation theory[51]. The most relevant bands are $\nu = -2$ (blue), $\nu = -1$ (orange), $\nu = 1$ (green), and $\nu = 2$ (red). The corresponding wave functions at the $K$ point for $\nu = -1$ and $\nu = 1$ are shown in Fig. 1**c**. Solid (dashed) lines represent the bands in the valley close to the $K$ ($K'$) point of the original graphene layers. The panel on the right shows the density of states with the colour shading representing band occupation at room temperature. Vertical arrows highlight the most relevant interband optical transitions. **b**. Calculated real part of the optical conductivity (black thick line) using the Kubo formula and the band structure in panel **a**. Blue (orange) shading represents the contribution to the total optical conductivity of the pair of bands with $\nu = -2, \nu' = 1$ ($\nu = -1, \nu' = 2$), corresponding to the transition marked by a blue (orange) arrow in panel **a**. The thin dashed line is the Lorentzian fit to the most relevant interband feature and is used to extract the resonance parameters. The inset shows the imaginary part of the optical conductivity (normalized to $G_0$), together with the experimental data from Fig. 3**c**. **c-d**. Same as in panels **a-b** but with the band structure of the chirally-symmetric continuum model[20].

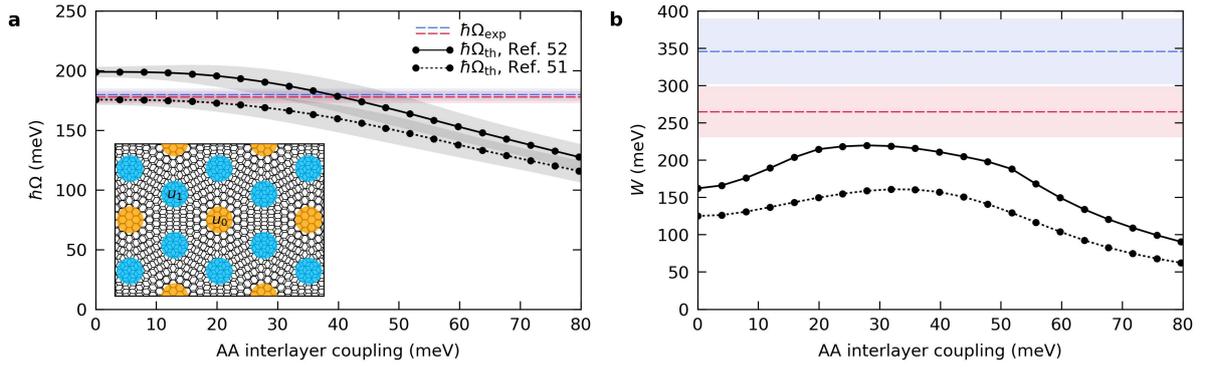

**Figure 5 | Calculated properties of the relevant interband transition as functions of the tunnelling amplitude in the AA regions. a.** Interband energy extracted from a Lorentzian fit of the optical conductivity. The grey shaded areas represent plus/minus the half width at half the maximum of the Lorentzian fit. In both theoretical models used in this work (Ref. 51 and 52), $\Omega_{th}$ decreases monotonically upon increasing the AA interlayer coupling $u_0$. These calculations were performed by setting $\theta = 1.35°$ and $u_1 = 97.5$ meV. The blue and red dashed lines correspond to the experimentally-determined resonant frequencies (Fig. 3c), with the corresponding shaded area indicating the uncertainty. The inset illustrates a triangular moiré lattice with the interlayer coupling strengths $u_0$, $u_1$ on the AA (yellow) and AB/BA sites (blue), respectively. **b.** Same as in **a** but for the spectral weight $W_{th}$. The latter displays a maximum for intermediate values of $u_0$.

# Supplementary Information for "Collective electronic excitations in twisted bilayer graphene close to the magic angle"

## Contents

# 1. Fabrication methods and list of studied devices

In this Section, we describe the fabrication of the device discussed in the main text, which has a twist angle of $1.35°$. The fabrication of the other devices follows similarly. We assemble the van der Waals heterostructure via the standard tear-and-stack dry transfer technique using a polycarbonate film (PC) (Fluka Analytical, #181641). This approach allows fabricating high-quality devices with clean interfaces between the different 2D materials and offers a high control of the twist angle[40,41]. We obtain all the employed 2D materials via micromechanical exfoliation onto a $SiO_2$/Si chip, and carefully select high-quality flakes using optical microscopy and atomic force microscopy.

The first step is to define the bottom gate on a Si/$SiO_2$ substrate using electron-beam lithography (EBL) and thermal evaporation of Cr and Pd/Au, resulting in a total thickness of 37 nm. Heat annealing in forming gas ($H_2$:Ar) is employed to clean the gate for 4 hours at $300°C$. Then, we place the bottom hBN flake on top of the metallic gate and anneal the structure once more to remove possible traces of residues from the stacking process. We inspect the structure after each annealing step using optical microscopy and and atomic force microscopy in order to confirm that the gate and hBN are flat and clean. After that, a heated PC film picks up the top hBN sheet, which picks up a portion of the graphene flake while tearing it. The remaining graphene is picked up after twisting the stage slightly more than the target angle to account for some relaxation of the twist angle. Finally, we deposit the hBN/twisted bilayer stack on top of the pre-made hBN/gate structure and select the twisted bilayer graphene regions with the fewest bubbles for further fabrication.

In order to pattern the stack into a Hall bar, we first define an etching mask using EBL (PMMA 950-A5 from Microchem, thickness $250$ nm), followed by reactive ion etching in an oxygen, argon, and $CHF_3$ atmosphere. Then, one dimensional Cr/Au edge-contacts to the TBG are defined via EBL (double resist recipe PMMA 495-A5/950-A2), reactive ion etching and thermal evaporation[S1]. Figure S1a shows an optical picture of our main device.

Table S1 outlines an overview of all the studied devices. The transport characteristics are reported where available. We have found the optically active regions in 5 devices (marked in green), including one made of twisted double bilayer graphene. The twist angle and CNP are determined using the method outlined in Section 2 of this file.

**Table S1 | Overview of the analysed devices in this work.**

| Device # | Device structure | Twist angle | Doping when ungated | Remarks |
|---|---|---|---|---|
| 1 | hBN-encapsulated TBG on metal gate | $1.35° \pm 0.02°$ | $\approx -5 \cdot 10^{11}$ cm$^{-2}$ at LT and RT. | Modes only observed after AFM-brooming. Defected contacts/gate. |
| 2 | | 1.5° | $\approx -7 \cdot 10^{11}$ cm$^{-2}$ at LT. $\approx -1.3 \cdot 10^{12}$ cm$^{-2}$ at RT. | Modes observed without AFM-brooming. Defected contacts/gate. |
| 3A[†] | | | | After AFM-brooming, modes appeared between s-SNOM measurements. |
| 3B[†] | hBN-encapsulated twisted double bilayer graphene on metal gate | | | Modes disappeared after AFM-brooming for a second time. |
| 4 | hBN-encapsulated TBG on metal gate | $1.10° \pm 0.05°$ | $\approx -7 \cdot 10^{11}$ cm$^{-2}$ at LT. | Modes observed without AFM-brooming, but not in 4-probe area. Device is superconducting. |
| 5A[†] | | (assume value Dev. 5B) | (assume value Dev. 5B) | Contacts slowly degrading over time. |
| 5B[†] | | 1.7° | $\approx 1 \cdot 10^{11}$ cm$^{-2}$ at LT. $\approx -2.4 \cdot 10^{12}$ cm$^{-2}$ at RT. | Defected contacts/gate. |
| 6 | | | $\approx 0$ cm$^{-2}$ at RT. | Gate does not affect twisted bilayer region (but does on single-layer graphene). |
| 7 | | | $\approx -2 \cdot 10^{11}$ cm$^{-2}$ at RT. | |
| 8 | hBN-encapsulated twisted double bilayer graphene on metal gate | $1.08° \pm 0.01°$ | $\approx 8.8 \cdot 10^{10}$ cm$^{-2}$ at LT. $\approx -1.0 \cdot 10^{12}$ cm$^{-2}$ at RT. | |
| 9, 10 | hBN-encapsulated TBG | | | |
| 11-13 | WSe$_2$-encapsulated TBG | | | |

[†]These pairs of devices are fabricated from the same heterostructure, and thus we assume their twist angle is similar.

## 2. Low-temperature transport characterization to determine the twist angle

We characterize the transport properties of our devices in a dilution refrigerator, with a base temperature of $\approx 70$ mK, which is equipped with a superconducting magnet generating a magnetic field perpendicular to the TBG electron gas. All the transport data are acquired using standard low-frequency lock-in techniques with discrete and distributed cryogenic low-pass filters removing thermal noise from the biasing and measurement lines[S2]. We bias the device with a fixed current of 10 nA and measure the pre-amplified four-probe voltages using SR-830 lock-in amplifiers that were synchronized to a frequency in the range $1 - 20$ Hz.

We extract the twist angle and CNP from cryogenic magnetotransport measurements (see Fig. S1b-c)[9,40]. For small twist angles $1° < \theta < 3°$, the bandgaps between the nearly-flat bands and the nearest conduction and valence bands cause strongly insulating states in transport measurements at characteristic carrier densities of $\pm n_s$[5,40,S3]. This density corresponds to the inverse of the superlattice unit cell, and by taking the double spin and valley degeneracy into account we determine the twist angle as

$$\theta[\text{rad}] \approx \sqrt{\frac{\sqrt{3}d_0^2}{8}n_s},$$

with $d_0$ being the lattice constant of graphene. We determine $n_s$ by extrapolating the Landau levels measured at high magnetic fields around these insulating states, down to zero magnetic field (see Fig. S1b). This yields $\theta = 1.35°$ with an uncertainty of $0.02°$. As part of this procedure, we calibrate the carrier density to the applied gate voltage by fitting the slope of the Landau levels around the CNP, which appear at $n = \nu B/\phi_0$, with filling factors $\nu = \pm 4, 8, 12 ...$, where $\phi_0 = h/e$ is the magnetic flux quantum. As a cross-check we also extracted the electronic density from Hall measurement near the CNP. As shown in Fig. S1c, this second procedure yields a result that agrees well with the density obtained from the slopes of the Landau levels.

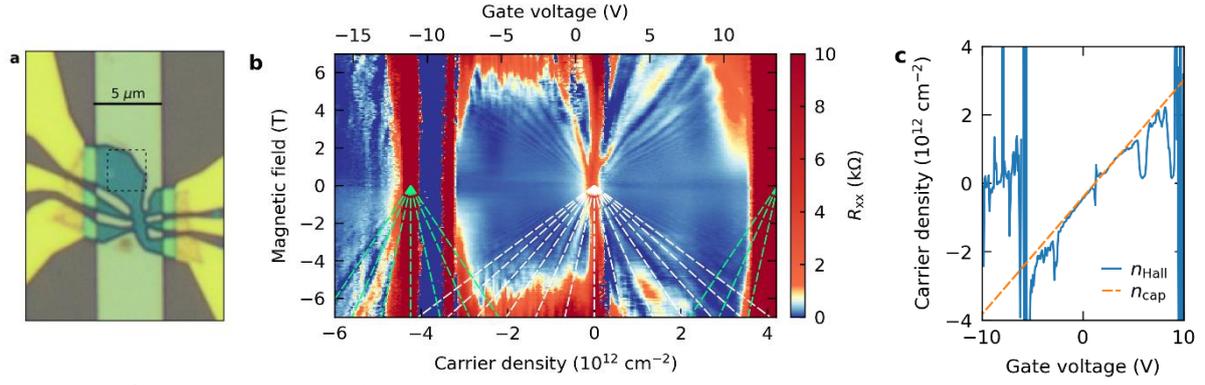

**Figure S1 | Determination of the twist angle using cryogenic transport measurements. a**. Optical picture of Device 1 shaped into a Hall bar on top of a metallic gate. The dashed area outlines the area where we performed s-SNOM measurements **b**. Longitudinal resistance showing the Landau levels originating from the CNP (white dashed lines) and from the insulating states at full filling (green dashed lines, centred at $n_s = \pm 4.24 \cdot 10^{12}$ cm$^{-2}$). All the transport measurements were performed at $\approx$ 70 mK. **c**. The carrier density $n_{\text{Hall}} = \frac{B}{-eR_{xy}}$ measured from the Hall resistance $R_{xy}$ at a magnetic field $B = 1$ T (blue curve), which agrees well with the carrier density $n_{\text{cap}}$ calculated from the slopes of the Landau levels (orange curve). The extracted capacity between the sample and the bottom gate, taking into account the measured bottom hBN thickness of 47 nm, corresponds to an out-of-plane static dielectric constant of the hBN $\epsilon_\perp(0) = 2.91$.

## 3. s-SNOM measurement setup

The near-field measurements presented in this work are carried out with a commercial scattering-type scanning near-field optical microscope (s-SNOM) (Neaspec). A tuneable quantum cascade laser (Daylight Solutions) and a $CO_2$ gas laser (Access Laser) are used as mid-infrared light sources with a typical power of $5-10$ mW, and the light is focussed on the metal-coated AFM tip (Nanoworld). The AFM tip oscillates at a frequency $\approx 250$ kHz with a tapping amplitude of $80-100$ nm, and modulates the scattered optical signal that is measured with a fast cryogenic HgCdTe detector (Kolmar Technologies). The near-field contribution to the optical signal is strongly dependent on the tip-sample distance, and therefore we can filter out the far-field contributions by locking to the 3$^{rd}$ harmonic.

We operate the s-SNOM in a non-interferometric detection scheme to obtain the optical signal $S_{\text{opt}}(x)$, as presented in the main text. This offers a higher signal to noise ratio compared to pseudoheterodyne detection, and allows us to vary the excitation wavelength while scanning. This non-interferometric scheme can cause small far-field contributions to the measured signal, which we have verified to be irrelevant for our conclusions. We performed a small planar correction to the data to correct for signal loss during the scans.

To ensure clean and flat surfaces for the s-SNOM measurements, we perform AFM-brooming after the low-temperature transport characterization. Repeatedly scanning in contact-mode over the sample removes from the sample surface residues originating from the fabrication process. These residues mask the fine optical and topographic features studied in this work, and form therefore a hurdle for s-SNOM measurements. Additionally, this AFM-brooming procedure helps to minimise the residues picked up by the apex of the AFM tip, and therefore reduces the loss of near-field signal during s-SNOM measurements. For this cleaning procedure, we use a standard AFM (Veeco) equipped with soft tips (Veeco OTR8-35, stiffness 0.15 N/m, and Nanoworld Arrow CONT, stiffness 0.2 N/m) and scan over the surface with a typical force $20-40$ nN and speed $2-4$ µm/s for several hours. After this procedure, the root-mean-square roughness is $120-200$ nm, which is similar to that of a pristine hBN flake[S4].

# 4. Correlation between optical activity, topography and force applied by the AFM tip

In some cases, we found a correlation between the optically active regions of our samples and fine features in the sample topography, as measured simultaneously using atomic force microscopy. Fig. S2 provides examples of this correlation and shows the presence of height steps of about 4 Å. These coincide with the boundaries of the optically active regions, which correspond to the higher areas. In addition, the small bubbles of a few nanometres thick, which are common for these van der Waals heterostructures, coincide in some cases with the boundaries of the optically active regions, or are located at the centre of regions where no collective excitations were measured. In addition, we note that in the core of device 4 (the area surrounded by 4 probes) we did not observe the collective modes, while being superconducting at low temperatures and not having been touched by contact-mode AFM. This illustrates that being close to magic angle itself is not sufficient for the formation of interband plasmons and that strain might be an additional requirement, or that excitations are outside our laser wavelength window.

We have observed areas of collective excitations appearing and disappearing in response to the tapping- and contact-mode atomic force microscopy measurements (Fig. S3). These changes go along with small, but critical changes in the sample topography. Although it is difficult to deduce precisely which local parameters are changing (i.e. twist angle, interlayer distance and coupling), it demonstrates that this system is rather sensitive to external forces[S5].

These observations point to the crucial role strain plays in the formation of these regions. This goes along with variations of the twist angle and interlayer coupling as commonly found in TBG samples[9,19,21,22], and directly affects the properties of the plasmonic modes.

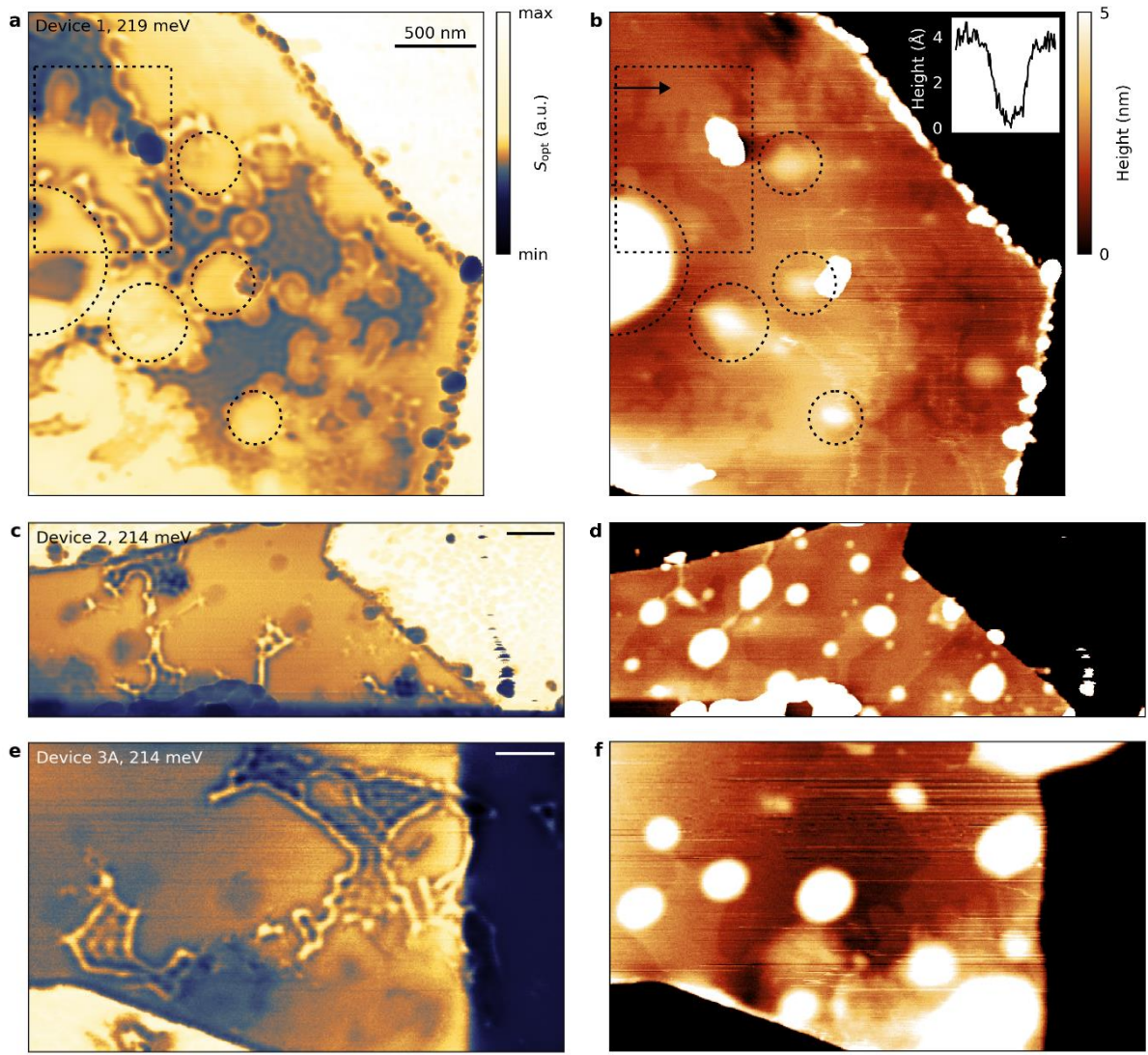

**Figure S2 | Correlation between optical activity and topography. a**. Near-field image of Device 1 as shown in the main text, with the corresponding AFM data shown in panel **b**. The rectangular area (dashed line) encloses several boundaries of optically active regions, which have a height step of several Å (inset shows line cut along the black arrow). The dashed circles mark small bubbles within the 2D heterostructure surrounded by regions with collective excitations. **c-f**. Same as in **a-b** but for Device 2 and 3A. All scale bars are 500 nm.

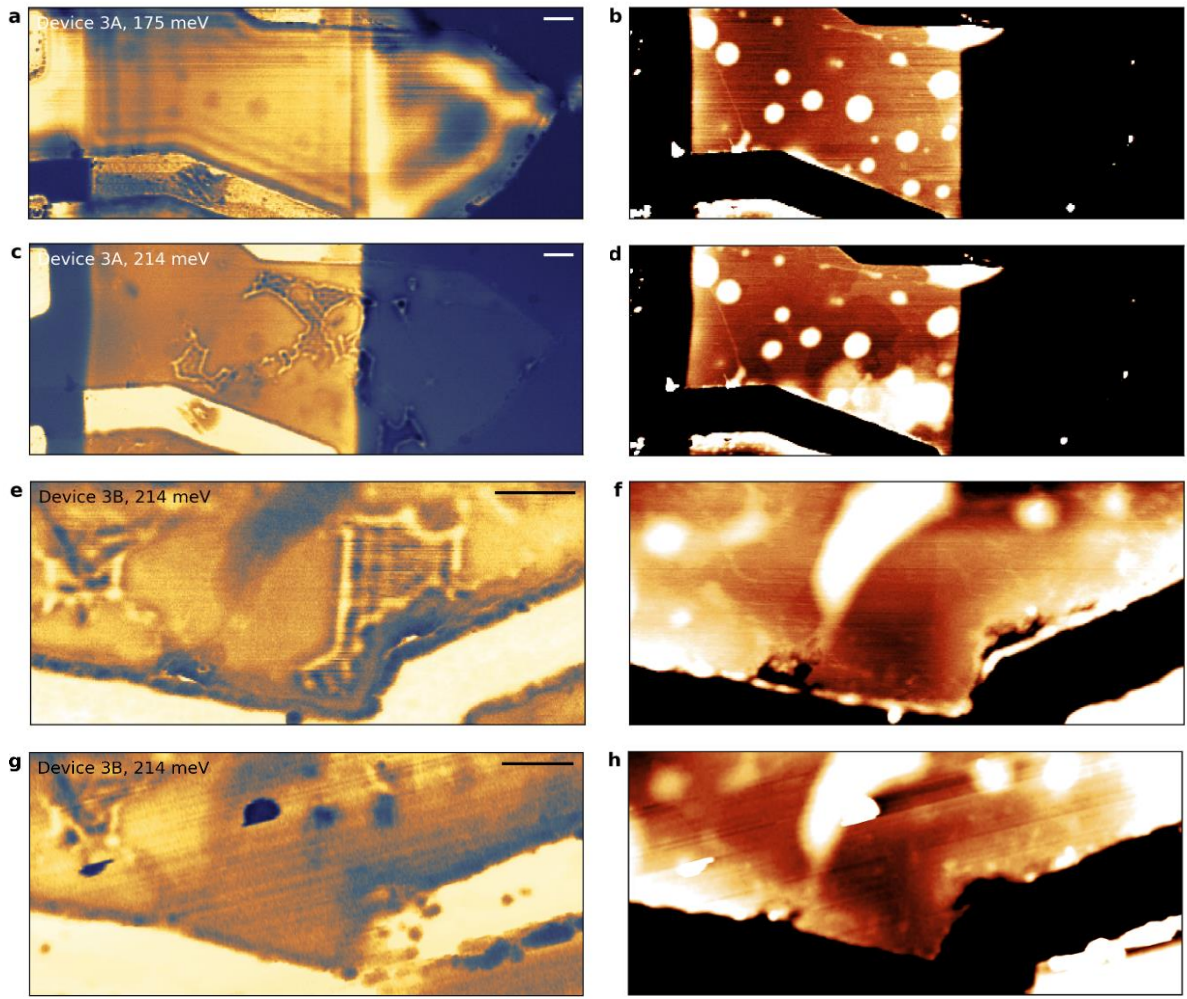

**Figure S3 | Changes in optically active areas induced by AFM. a**. Near-field image and corresponding topography (panel **b**) before a structural change occurred during a s-SNOM measurement with the AFM operated in tapping-mode. The fringes parallel to the sample edges are hBN phonon-polaritons. After this change, the topography shows a few alterations and the collective excitations appear (**c-d**). **e-f**. Collective excitations in twisted double bilayer graphene before a second AFM-brooming session, which disappeared afterwards (**g-h**). All scale bars are 500 nm and the colormaps are the same as in Fig. S2 (except in panels **f** and **h,** where it spans 10 nm).

# 5. Extraction of the wavevector from the experiment

To determine the plasmon dispersion $q(\omega) = q_1(\omega) + iq_2(\omega)$ we measure one-dimensional linetraces while varying the incident photon energy $\hbar\omega$. We fit the observed oscillations in the optical signal with the function $S_{\text{opt}}(x) = \text{Re}[A\,x^{-1/2}\,e^{2iqx}] + Bx$. Here, $x$ is the tip position along the line cut, as measured from the interface, $A \equiv A_1 + iA_2$ and $q \equiv q_1 + iq_2$ are complex fit parameters, and $B$ represents a linear background[32]. In this interpretation, we assume that the plasmons are predominantly launched by the s-SNOM tip[S6] and reflected by the boundaries of the optically active regions. This requires the factor of two in the exponential function and the geometrical $1/\sqrt{x}$ decay. Because of the lack of topographic features along the presented line cuts, we define the position of the reflecting interface to be at the peak position of $S_{\text{opt}}(x)$. Fig. S4 presents the results of this fitting procedure for a few selected photon energies. We use the real part of the wavevector, $q_1$, to calculate the imaginary part of optical conductivity, while the error in $q_2$ is too large to make any precise statement about the lifetime of the collective excitations.

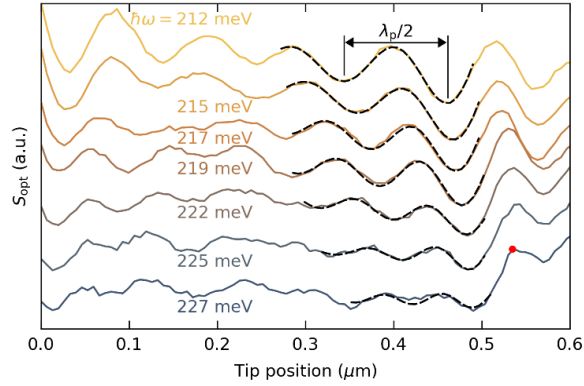

**Figure S4 | Fitting procedure to determine the interband plasmon wavevector.** Coloured lines are selected linetraces taken from Fig. 3**a** of the main text for various illumination energies. The lines are separated vertically for clarity. From a fit according to our model (dashed lines), we extract the plasmon wavevector $q_1 = 2\pi/\lambda_{\text{p}}$ as presented in Fig. 3**b** of the main text. The red dot marks the position of the reflecting interface used in the fits.

# 6. Comparison with plasmons in doped single-layer graphene

Single-layer graphene (SLG) can display plasmonic excitations[32], though there are key differences with the observations made here for TBG. To make a comparison, we calculate the loss function $L(q,\omega)$ as introduced in the main text for the two systems. For SLG we use the local, frequency dependent conductivity taken from Ref. S7, while for the conductivity of TBG we employ the chirally-symmetric continuum model of Ref. 20. We perform all calculations with a temperature of 300 K and consider the finite-thickness effect of hBN. The simulated device structure consists of a metal gate, 47 nm bottom hBN, SLG or TBG, and 13 nm top hBN. Figure S5 displays the results.

First and foremost, we find that only highly doped SLG can host plasmonic excitations with a dispersion qualitatively (but not quantitatively) similar to what we have found (Fig. S5b). For that case we considered a Fermi level $E_\mathrm{F} = 0.5$ eV, which can be reached via electrostatic gating only at the breakdown voltage of hBN ($\sim 1$ V/nm). The fact that we observe the collective excitations in ungated (thus nearly charge-neutral) TBG marks a strong difference from SLG.

In addition, the typical SLG plasmons are predominantly observed near the edges of the device or particular defects within the samples[45,46] The observation of enclosed patches of optical activity in TBG, formed by reflecting interfaces possibly linked to local strain and stress variations (see Sect. 4), is something not seen in SLG.

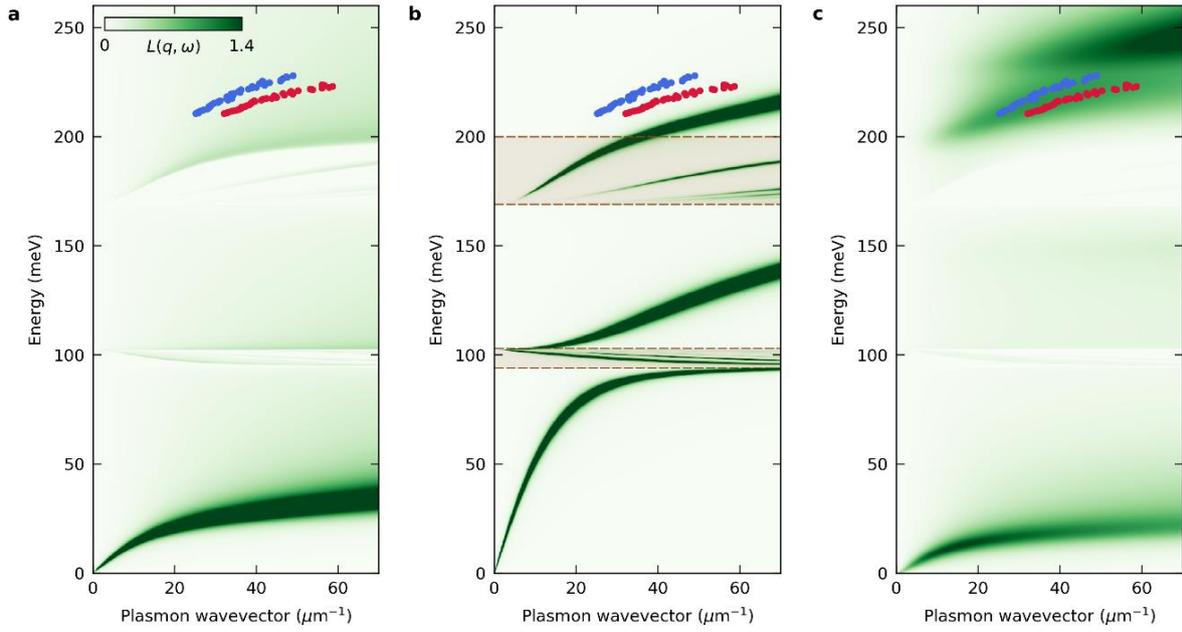

**Figure S5 | Comparison of calculated loss function for SLG and TBG. a**. The loss function of SLG at $E_\mathrm{F} = 0$. The strong excitations at low energies are graphene plasmons at THz-frequencies resulting from thermally excited carriers[S8]. The data points are the same as those in Fig. 2**b** of the main text. **b**. Same as in panel **a** but for $E_\mathrm{F} = 0.5$ eV, corresponding to a doping level of $1.85 \cdot 10^{13}$ cm$^{-2}$. The brown dashed areas indicate the lower and upper hBN reststrahlen bands. **c**. The loss function of TBG as calculated from the chirally-symmetric continuum model[20] for $\theta = 1.35°$.

# 7. Details of the extraction of optical conductivity

In the local approximation for the optical conductivity[48], the longitudinal dielectric function is given by

$$\epsilon(q,\omega) = 1 + \frac{iq^2 V_{q,\omega}}{\omega}\sigma(\omega).$$

Here, $\sigma(\omega)$ is the local, frequency-dependent conductivity and $V_{q,\omega}$ is the electron-electron (e-e) interaction potential. Note that $\sigma(\omega)$ is a scalar because the system has a vertical $C_3$ symmetry axis. The e-e interaction potential relates a charge density fluctuation $\rho(\boldsymbol{q},\omega)$ in the electron gas to the electric potential $\phi(\boldsymbol{q},\omega)$ it induces through

$$\phi(\boldsymbol{q},\omega) = V_{q,\omega}\,\rho(\boldsymbol{q},\omega).$$

For the type of structure used in our experiments, the e-e interaction potential can be calculated following Ref. S8. This yields

$$V_{q,\omega} = \frac{2\pi}{q\,\tilde{\epsilon}(\omega)} F(q,\omega),$$

where $\tilde{\epsilon}(\omega) = \sqrt{\epsilon_\parallel(\omega)\epsilon_\perp(\omega)}$ is the average permittivity of hBN, $\eta(\omega) = \sqrt{\epsilon_\parallel(\omega)/\epsilon_\perp(\omega)}$ is its anisotropy factor, and

$$F(q,\omega) = \frac{\{[\tilde{\epsilon}(\omega)+1]+[\tilde{\epsilon}(\omega)-1]e^{-2q\eta(\omega)t_2}\}[1-e^{-2q\eta(\omega)t_1}]}{[\tilde{\epsilon}(\omega)+1]+[\tilde{\epsilon}(\omega)-1]e^{-2q\eta(\omega)(t_1+t_2)}},$$

$t_1$ and $t_2$ being the bottom and top hBN thickness, respectively. We remind the reader that hBN is an uniaxial crystal with the optical axis $\hat{\boldsymbol{z}}$ perpendicular to the plane of the flake. Its dielectric tensor is diagonal in the $\hat{\boldsymbol{x}}, \hat{\boldsymbol{y}}, \hat{\boldsymbol{z}}$ basis with $\epsilon_{xx}(\omega) = \epsilon_{yy}(\omega) = \epsilon_\parallel(\omega)$ and $\epsilon_{zz}(\omega) = \epsilon_\perp(\omega)$. Each component is described by an oscillator model

$$\epsilon_i(\omega) = \epsilon_i(\infty) + \frac{s_i \hbar^2 \omega_i^2}{\hbar^2 \omega_i^2 - i\hbar^2 \gamma_i \omega - \hbar^2 \omega^2},$$

with parameters given in Ref. S9 and reported for completeness in Table S2.

The mode penetrates for a characteristic length $\ell_z = [q\eta(\omega)]^{-1}$ in the hBN slab, in the vertical direction. To couple efficiently to the s-SNOM tip, the thickness of the top hBN layer $t_2$ has to be smaller than $\ell_z$. Note that close to the upper edge of the higher hBN reststrahlen band, at $\hbar\omega \approx 200$ meV, $\ell_z$ diverges due to the vanishing of $\epsilon_\parallel(\omega)$.

Since the procedure we have used to extract the optical conductivity relies on the knowledge of the e-e interaction potential, it is important to quantify how uncertainties in this quantity propagate through, affecting the final results. One source of uncertainty is the model used for the hBN permittivity. We compared different models found in literature (see e.g. Refs. S9, S10, and S11) and found an agreement within 3% with the model we used, within the frequency range where experimental data points lie. The second, and most important source of error, is the uncertainty in the thicknesses of the hBN crystal slabs that enter the expression of $F(q,\omega)$. The hBN thickness is measured by AFM. For the device shown in the main text, we find $t_1 = 47 \pm 5$ nm and $t_2 = 13 \pm 3$ nm. To estimate the error we repeated our analysis spanning the two thicknesses inside these confidence intervals (see Fig. S6) and used the maximum deviation of the obtained parameters from the central values as a measure of the error introduced on $\hbar\Omega_\text{exp}$ and $W_\text{exp}$. Since this error is larger than the error coming from other sources, we used it as a measure of the error on the extracted parameters.

Note that the hBN thickness, although not known precisely, is uniform inside a given device and therefore the difference between the parameters extracted in different locations of the same sample is still meaningful even if smaller than the error.

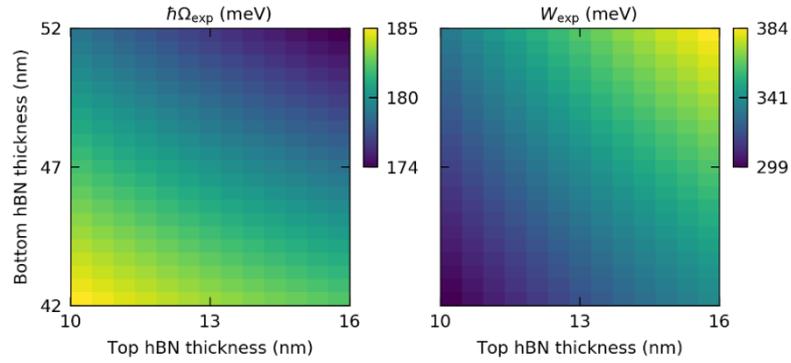

**Figure S6 | Propagation of the error due to hBN thickness uncertainty.** Extracted energy and spectral weight of the resonance as a function of the two hBN thicknesses within the confidence interval. Calculations refer to the blue data set in Fig. 3**b**-**c** of the main text. We performed the same analysis on the other data set**.**

**Table S2 | Parameters of the model of the hBN permittivity.**

|  | $\epsilon_i(\infty)$ | $s_i$ | $\hbar\omega_i$ (meV) | $\hbar\gamma_i$ (meV) |
|---|---|---|---|---|
| $i = \parallel$ | 4.9 | 2.001 | 168.6 | 0.87 |
| $i = \perp$ | 2.95 | 0.5262 | 94.2 | 0.25 |

# 8. Bands, optical conductivity, and loss function for different twist angles and interlayer couplings

In this section we provide additional plots of the band structure, optical conductivity, and loss function as calculated for different values of the AA interlayer tunnelling amplitude $u_0$ and twist angle $\theta$. Eigenvalues $\epsilon_{k\nu}$ and eigenfunctions $|k\nu\rangle$ of the single-particle Bloch problem are calculated according to the models in Ref. 51 and Ref. 52.

The optical conductivity is computed according to the Kubo formula

$$\sigma(\omega) = g\pi i G_0 \left[ \frac{1}{\hbar\omega} \sum_\nu \int_{BZ} \frac{d^2\boldsymbol{k}}{(2\pi)^2} (-f'_{\boldsymbol{k}\nu}) \left|\langle \boldsymbol{k}\nu | \frac{\hbar \hat{p}_x}{m_e} | \boldsymbol{k}\nu \rangle\right|^2 + \sum_{\nu\neq\mu} \int_{BZ} \frac{d^2\boldsymbol{k}}{(2\pi)^2} \left(-\frac{f_{\boldsymbol{k}\nu}-f_{\boldsymbol{k}\mu}}{\epsilon_{\boldsymbol{k}\nu}-\epsilon_{\boldsymbol{k}\mu}}\right) \frac{\left|\langle \boldsymbol{k}\nu | \frac{\hbar\hat{p}_x}{m_e} | \boldsymbol{k}\mu\rangle\right|^2}{\epsilon_{\boldsymbol{k}\nu}-\epsilon_{\boldsymbol{k}\mu}+\hbar\omega+i\eta} \right].$$

Here, $f_{\boldsymbol{k}\nu}$ is the Fermi distribution evaluated at the energy $\epsilon_{\boldsymbol{k}\nu}$, $f'_{\boldsymbol{k}\nu}$ is its energy derivative, $g = 4$ takes into account the valley and spin degeneracy, $\eta$ is a small positive number, $\hat{p}_x$ is the $x$ component of the canonical momentum operator, and $m_e$ is the bare electron mass in vacuum. The discrete sums run over the band indices $\nu$ and $\mu$, and the 2D integrals over the moiré superlattice Brillouin zone (BZ).

The matrix elements $\langle \boldsymbol{k}\nu | \frac{\hbar\hat{p}_x}{m_e} | \boldsymbol{k}\mu\rangle$ are calculated according to

$$\langle \boldsymbol{k}\nu | \frac{\hbar\hat{p}_x}{m_e} | \boldsymbol{k}\mu\rangle = \langle u_{\boldsymbol{k}\nu} | \frac{\hbar(\hat{p}_x + \hbar k_x)}{m_e} | u_{\boldsymbol{k}\mu}\rangle = \langle u_{\boldsymbol{k}\nu} | \frac{\partial H(\boldsymbol{k})}{\partial k_x} | u_{\boldsymbol{k}\mu}\rangle,$$

where $|u_{\boldsymbol{k}\mu}\rangle$ is the periodic part of the Bloch function corresponding to crystal momentum $\hbar\boldsymbol{k}$ and band index $\nu$. Wavevector space is sampled using a $60 \times 60$ uniform mesh of the parallelogrammatic BZ, corresponding to 641 points in the irreducible ($D_3$) BZ. We expanded the periodic parts of the wave functions on a plane wave basis, keeping wavevectors up to the 8[th] shell of the reciprocal lattice of the moiré superlattice. In the calculation of the optical conductivity we included the contributions of twenty bands above and twenty bands below the Dirac point of the original graphene layers and we set $\eta = 5$ meV.

Additional numerical results, with respect to those presented in the main text, are reported in Figs. S7, S8, and S9. In Fig. S7 we see that the resonance in the optical conductivity shifts towards higher frequencies, sharpens, and displays a larger spectral weight when $u_0$ is reduced from 79.7 meV to 19.9 meV. A further reduction of $u_0$ down to 0 meV leads to an additional blueshift and sharpening, but the spectral weight decreases, in agreement with Fig. 5 of the main text. This proves that good agreement with experimental data is possible for a range of values of $u_0$, that are smaller than the commonly accepted value 79.7 meV, but still finite. The same trends can be seen in Fig. S8 where the model in Ref. 51 is used to calculate the electronic band structure.

In Fig. S9 we instead visualize the impact of the twist angle $\theta$ on the results of our calculations. A small deviation (0.05°) from the experimentally measured $\theta = 1.35°$ has a very limited impact on the calculated optical properties (see comparison between Figs. S9d-f and Figs. 3b,4c-d of the main text). Upon increasing the twist angle, the resonance shifts monotonically to higher energies. We note that the resonance is sharpest and closest to the experimental data when the angle is close to the independently measured value $\theta = 1.35°$ despite the absence of any fitting parameter in the theory.

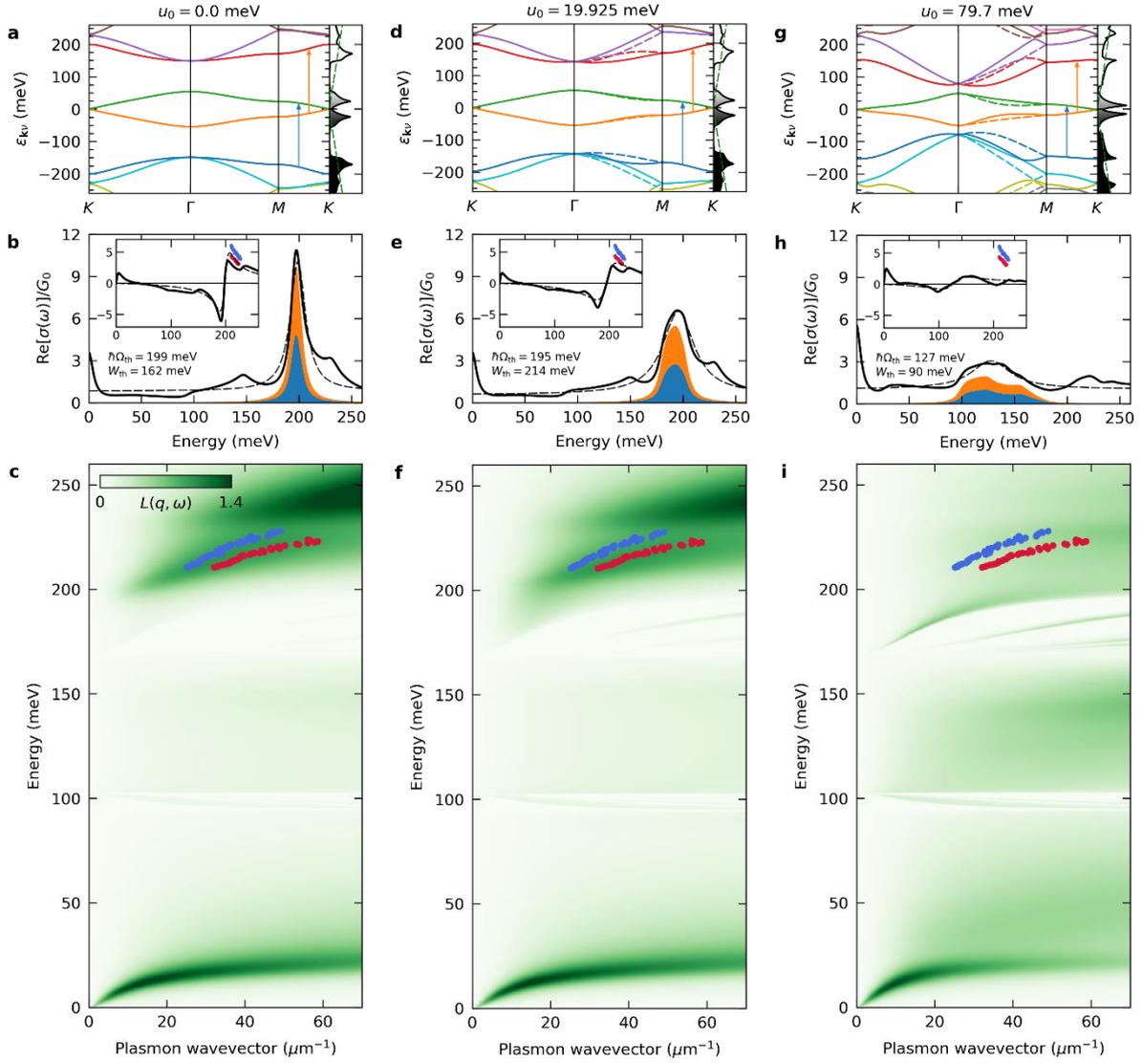

**Figure S7 | Impact of the AA interlayer tunnelling amplitude electronic bands, conductivity, and loss function, using model in Ref. 52. a.** Band structure of TBG with $\theta = 1.35°$ calculated according to the model in Ref. 52 with $u_0 = 0$. Color coding is the same as in Fig. 4**a** of the main text. **b.** Optical conductivity derived from the band structure in **a**. Color coding is the same as in Fig. 4**b** of the main text. **c.** Loss function derived from the band structure in **a**. Color coding is the same as in Fig. 3**b** of the main text. Panels **d-f** and **g-i** are the same as **a-c** with $u_0 = 19.925$ meV and $u_0 = 79.7$ meV respectively.

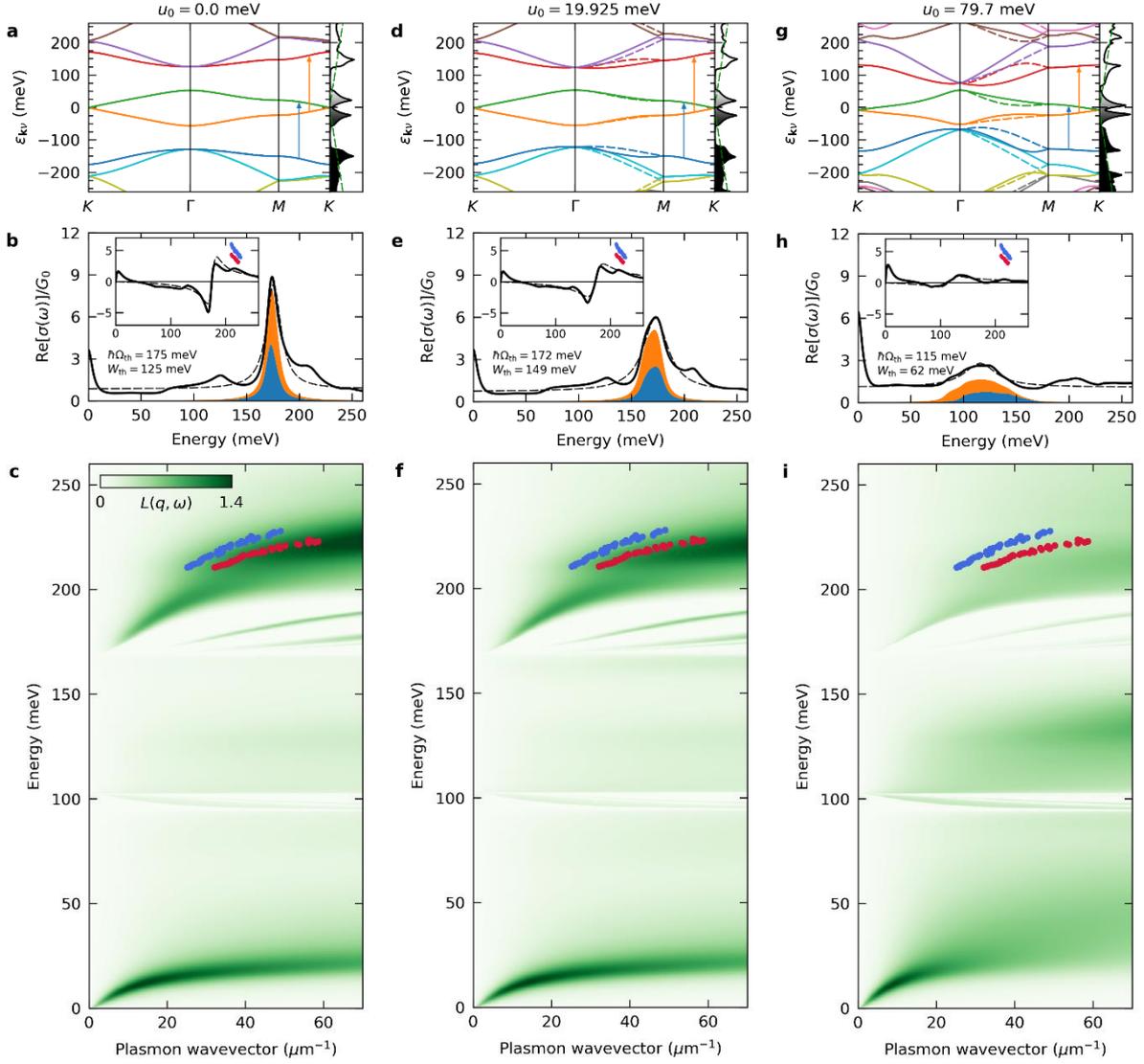

**Figure S8 | Impact of the AA interlayer tunnelling amplitude electronic bands, conductivity, and loss function, using model in Ref. 51. a**. Band structure of TBG with $\theta = 1.35°$ calculated according to the model in Ref. 51 with $u_0 = 0$ and $u_1 = 97.5$ meV. Color coding is the same as in Fig. 4**a** of the main text. **b**. Optical conductivity derived from the band structure in **a**. Color coding is the same as in Fig. 4**b** of the main text. **c**. Loss function derived from the band structure in **a**. Color coding is the same as in Fig. 3**b** of the main text. Panels **d**-**f** and **g**-**i** are the same as **a**-**c** with $u_0 = 19.925$ meV and $u_0 = 79.7$ meV respectively.

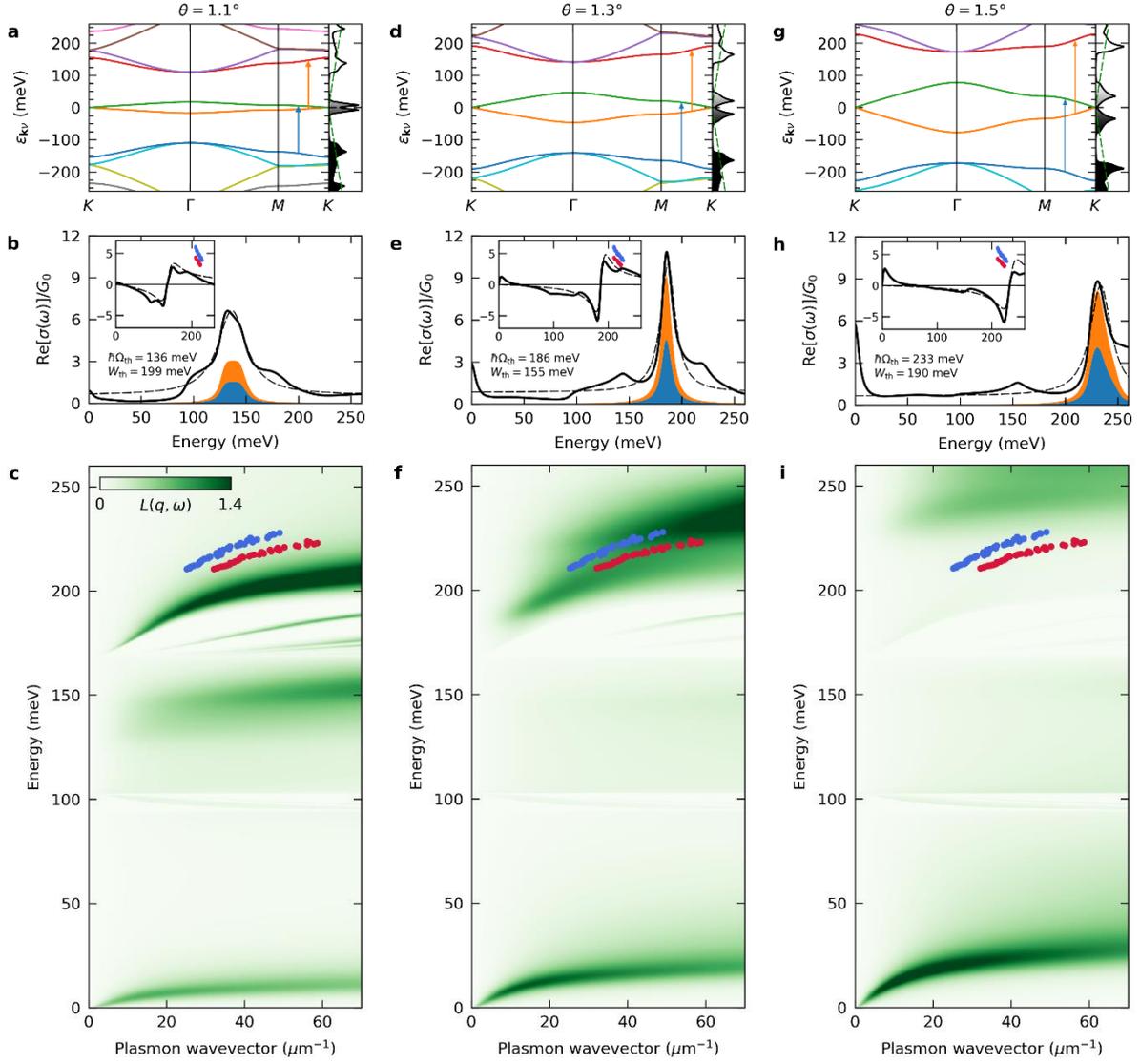

**Figure S9 | Impact of twisting angle on electronic bands, conductivity, and loss function. a.** Band structure of TBG calculated according to the model in Ref. 52 with $u_0 = 0$ (CS-CM) and $\theta = 1.1°$. Color coding is the same as in Fig. 4**a** of the main text. **b**. Optical conductivity derived from the band structure in **a**. Color coding is the same as in Fig. 4**b** of the main text. **c**. Loss function derived from the band structure in **a**. Color coding is the same as in Fig. 3**b** of the main text. Panels **d**-**f** and **g**-**i** are the same as **a**-**c** with for $\theta = 1.3°$ and for $\theta = 1.5°$ respectively.

# 9. Effects of a pseudo-magnetic field due to strain

In this Section, we investigate the hypothesis that our samples exhibit a highly inhomogeneous strain distribution with patches where the associated pseudomagnetic field $\boldsymbol{B}_S = B_S \hat{\boldsymbol{z}}$ is finite and nearly uniform[54] and regions where $\boldsymbol{B}_S = \boldsymbol{0}$. In the former ones, corresponding to the patches of optical activity, a bulk pseudo-magnetoplasmon mode could exist. Its long-wavelength gap can be estimated by the usual cyclotron frequency formula for single-layer graphene $\hbar\Omega_c = \sqrt{2}\,\hbar v_F/\ell_{B_S}$ or by its equivalent for Bernal-stacked bilayer graphene $\hbar\Omega_c = \hbar^2/(m^*\ell_{B_S}^2)$, where $\ell_{B_S} \approx 25$ nm/$\sqrt{B_S[\text{Tesla}]}$ is the magnetic length. Here, $v_F \approx 10^6$ m/s is the Fermi velocity in single-layer graphene and $m^* \approx 0.03\,m_e$ is the effective mass in Bernal-stacked bilayer graphene, $m_e$ being the electron mass. In order to find a long-wavelength gap on the order of 180 meV, one needs a pseudomagnetic field $B_S \approx 20 - 50$ T.

This field is one order of magnitude smaller than the pseudomagnetic field measured in highly-strained graphene nano-bubbles[S12]. However, due to the larger size of the regions of the sample where we observe optical activity, the amount of strain required to generate such a field is much larger. We can estimate it as follows.

Under triangular strain configurations it is possible to achieve a nearly-uniform pseudomagnetic field[54], with the maximum field at the centre of the strained area given by

$$B_S = \frac{8\beta\Delta_m}{Dd_0} \cdot \frac{\hbar}{e}.$$

Here $\Delta_m$ is the maximum strain, $D$ is the diameter of the area, $\beta \equiv -\frac{\partial \ln(t)}{\partial \ln(d_0)} \approx 2$, with $t$ being the nearest-neighbour hopping parameter and $d_0 = 0.246$ nm the graphene lattice constant. Figure S10 shows $B_S$ for strained areas of various diameters up to rather high amounts of strain. The patches with collective excitations occur in sizes up to 1 $\mu$m, and thus require unrealistically high strain values and we therefore conclude that this explanation of our observations is unlikely.

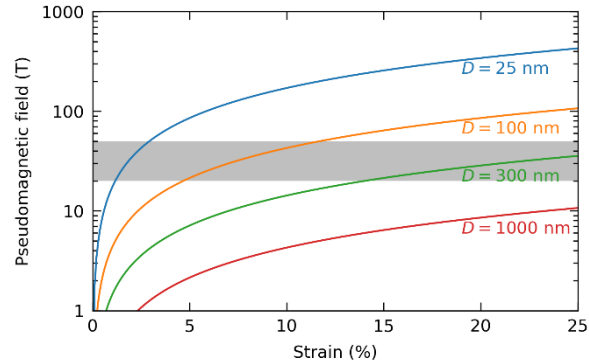

**Figure S10 | Pseudomagnetic field generated by strain.** The coloured lines display the maximum field generate at the centre of a strained area of single-layer graphene[54]. The shaded grey area marks the range of fields required to generate a pseudo-magnetoplasmon with a resonant interband energy around 180 meV.

# 10. Supplementary references